\documentclass[conference]{IEEEtran}
\IEEEoverridecommandlockouts
\usepackage{cite}
\usepackage{amsmath,amssymb,amsfonts}
\usepackage{algorithmic}
\usepackage[dvipdfmx]{graphicx}
\usepackage{textcomp}
\usepackage{url}
\usepackage{xcolor}

\ifCLASSOPTIONcompsoc
  \usepackage[caption=false,font=normalsize,labelfont=sf,textfont=sf]{subfig}
\else
  \usepackage[caption=false,font=footnotesize]{subfig}
\fi

\def\BibTeX{{\rm B\kern-.05em{\sc i\kern-.025em b}\kern-.08em
    T\kern-.1667em\lower.7ex\hbox{E}\kern-.125emX}}

\begin{document}

\title{Analysis of the Influence of Internet TV Station on Wikipedia Page Views}

\author{
    \IEEEauthorblockN{Hiroshi Hayano\IEEEauthorrefmark{1} \quad Masanori Takano\IEEEauthorrefmark{2} \quad Soichiro Morishita\IEEEauthorrefmark{2} \quad Mitsuo Yoshida\IEEEauthorrefmark{1} \quad Kyoji Umemura\IEEEauthorrefmark{1}}
    \IEEEauthorblockA{\IEEEauthorrefmark{1}\textit{Toyohashi University of Technology}\\
    Aichi, Japan\\
    h173351@edu.tut.ac.jp, yoshida@cs.tut.ac.jp, umemura@tut.jp}
    \IEEEauthorblockA{\IEEEauthorrefmark{2}\textit{CyberAgent, Inc.}\\
    Tokyo, Japan\\
    \{takano\_masanori, morishita\_soichiro\}@cyberagent.co.jp}
}

\maketitle

\begin{abstract}
We aim to investigate the influence of television on the web; 
if the influence is strong, a viral effect may be expected.
In this paper, we focus on the Internet TV station and on Wikipedia use as exploratory behavior on the web. 
We analyzed the influence of Internet TV station on Wikipedia page views.
Our aim is to clarify the characteristics of page views as related to Internet TV station in order to index outward impact and develop a prediction model.
The results indicate that there is a correlation between TV viewership and page views.
Moreover we find that the time lag between TV and web gradually reduce as broadcasts begin after 9:00; 
after 23:00, page views tend to be maximized during the broadcast itself.
We also differentiate between page views on  PC and on mobile 
and find that PC pages tend to be accessed more during the daytime.
In addition, we consider the number of broadcasts per program, 
and observe that viewership tends to stabilize as the number of broadcasts increases but that page views tend to decrease.
\end{abstract}

\begin{IEEEkeywords}
Page views, TV viewership, AbemaTV, Wikipedia
\end{IEEEkeywords}

\section{Introduction}

The most widely used index in the rating of television programs is viewership; 
increasing viewership generally leads to an increase in revenue from advertising.
However, there are limits to this index closed only in the TV.
It is well known that watching television can influence people's behavior~\cite{RUBIN1987,BORZEKOWSKI2001,Druckman2003}
but this impact cannot be measured by viewership figures.
In reverse, some previous studies have aimed to predict TV viewership from user behavior~\cite{Cheng2013,Fukushima2016}.
We wanted to understand the influence of television on the web; 
if the influence is strong, a viral effect may be expected.
Focusing on this topic, one previous study analyzed YouTube and Twitter~\cite{Abisheva2014}.
However, although YouTube has channels, it does not broadcast programs based on a schedule like television.

In this paper, we take Internet TV station as the television  element and browsing Wikipedia as an exploratory behavior on the web, 
and we analyze the influence of  Internet TV station on Wikipedia page views.
Our aim is to clarify the characteristics of page views as related to Internet TV station in order to index outward impact and develop its prediction model.
To this end, we address the following research questions:
\begin{description}
 \item[RQ1] Correlation between viewership and page views:\\
   Is there a correlation between viewership and page views?
 \item[RQ2] Time until page view is maximized:\\
   How long does it take for the page view to achieve its maximum after the program starts or ends?
 \item[RQ3] Maximum page view by time of day:\\
   When is the hour to achieve the maximum page view?
 \item[RQ4] Viewership and page views by number of broadcasts:\\
   Is there a difference in viewership and page views depending on the number of broadcasts?
\end{description}

The results indicated that there is a correlation between viewership and page views.
We also found that the time lag between TV and web  gradually reduces as broadcast start times move beyond 9:00;
after 23:00, page views tend to be maximized during the broadcast itself.
Moreover, we differentiate between page views on PC and on mobile, 
and find that PC pages tend to be accessed more during the daytime.
We also consider the number of broadcasts per program, 
and observe that viewership tends to stabilize as the number of broadcasts increases but that page views tend to decrease.

\section{Datasets}

\subsection{AbemaTV: Programs and Viewership}

In this study, we used the viewing history data of AbemaTV\footnote{\url{https://abema.tv/}} from June to October 2017.
AbemaTV is a well-known channel Internet TV station service in Japan provided by AbemaTV Inc.\footnote{\url{http://abematv.co.jp/}} which  broadcasts programs based on a schedule.

The AbemaTV data includes program ID and viewing start and end times for each user.
We used this data to calculate the number of users (\textit{i.e.}, the viewership) per program on an hourly basis with the following conditions:
\begin{itemize}
  \item We count only user who view programs continuously for more than 60 seconds.
  Viewing for less than 60 seconds is regarded as zapping and does not constitute viewership.
  \item We count only users who view programs in real time based on the schedule.
  Data on time shifting is excluded.
  \item We count for each viewing tool that is used:
  web services, native applications (for Android and iOS), and TV applications (for Apple TV, Google Cast, and Fire TV).
  \item We count only unique users;
  if the same user appears multiple times in one program, this counts as one.
\end{itemize}
Since viewing history data is data of the sampled users, the value of viewership in the paper is different from the actual value.

\subsection{Wikipedia: Articles and Page Views}

In this study, we used the page view data of Wikipedia\footnote{\url{https://dumps.wikimedia.org/other/pageviews/readme.html}} from May to November 2017.
By using this data, we were able to see how often certain topics were examined by Wikipedia users.
In addition, Wikipedia page views generally reflect the degree of public interest in a particular subject 
because they tend to correlate with web search frequency~\cite{Yoshida2015}.

The page view data includes article ID, and PC and mobile page views for each hour.
The data filters out as many spiders and bots as Wikimedia Foundation, Inc. can detect\footnote{\url{https://wikitech.wikimedia.org/wiki/Analytics/Data_Lake/Traffic/Pageviews}},
but the page view is not the unique number of users, and it is counted each time when the same user views multiple times in the article.

\subsection{Association of Programs and Articles}

We associated AbemaTV programs with Wikipedia articles 
using both Google and Wikipedia's internal search engine.
A program could be associated with an article according the following.
\begin{itemize}
  \item Articles about programs.
  \item Articles about works (mainly animation, drama, and movies).
  \item Articles about tournament name (mainly shogi and mahjong).
  \item Articles about tournaments (mainly sports).
\end{itemize}

A breakdown by genre of programs associated with articles is shown in TABLE~\ref{tab:genre}.
Here, ``Genre'' is the genre on AbemaTV,
``Programs'' is the number of programs that were able to associate with Wikipedia articles,
``Articles'' is the number of Wikipedia articles associated with the program,
and ``Rate'' is the rate at which programs can be associated with articles.
``Programs'' and ``Articles'' do not match because one article may be associated with multiple programs.
Conversely, in the case of an animation series composed of 12 episodes broadcast consecutively,
the number of programs is 12 but the series is only associated with one article.
As a result, 41,801 of 78,082 (53.5\%) programs on AbemaTV were associated with 588 articles on Wikipedia.
There are series composed of programs with multiple genres.
There were ten such series, and the sum of articles does not match ``Total''.

\begin{table}[tp]
  \centering
  \caption{Genres of associated programs: Most of the programs of animation, drama, and sports were associated, but the program of hobby and music was hardly associated.}
\begin{tabular}{ l | r r | r }
\hline
Genre & Programs & Articles & Rate \\
\hline
Animation & 27,477 & 325 & 0.982 \\
Asia & 2,657 & 39 & 0.430 \\
Documentary & 204 & 6 & 0.044 \\
Drama & 3,948 & 54 & 0.955 \\
Fighting sports & 569 & 6 & 0.296 \\
Fishing & 335 & 1 & 0.097 \\
Golf & 248 & 20 & 0.086 \\
Hip hop & 755 & 3 & 0.241 \\
Hobby & 185 & 1 & 0.033 \\
Mahjong & 166 & 8 & 0.193 \\
Movie & 95 & 47 & 0.833 \\
Music & 31 & 9 & 0.015\\
News & 397 & 5 & 0.187 \\
Reality & 1,142 & 3 & 0.265 \\
Shogi & 590 & 10 & 0.573 \\
Soccer & 661 & 8 & 0.525 \\
Sports & 329 & 13 & 0.820 \\
Variety & 2,012 & 40 & 0.447 \\
\hline
Total & 41,801 & 598 & 0.535 \\
\hline
\end{tabular}
  \label{tab:genre}
\end{table}

\section{Results and Discussion}

\subsection{Correlation between viewership and page views}

We analyze whether viewership (VS) and page views (PV) are correlated or not.
In other words, we conducted a basic analysis of a second screen indicating if users use the online search services while watching the Internet TV station.

\begin{figure*}[tp]
  \centerline{
  \subfloat[One episode of one program.]{
    \includegraphics[width=0.48\linewidth]{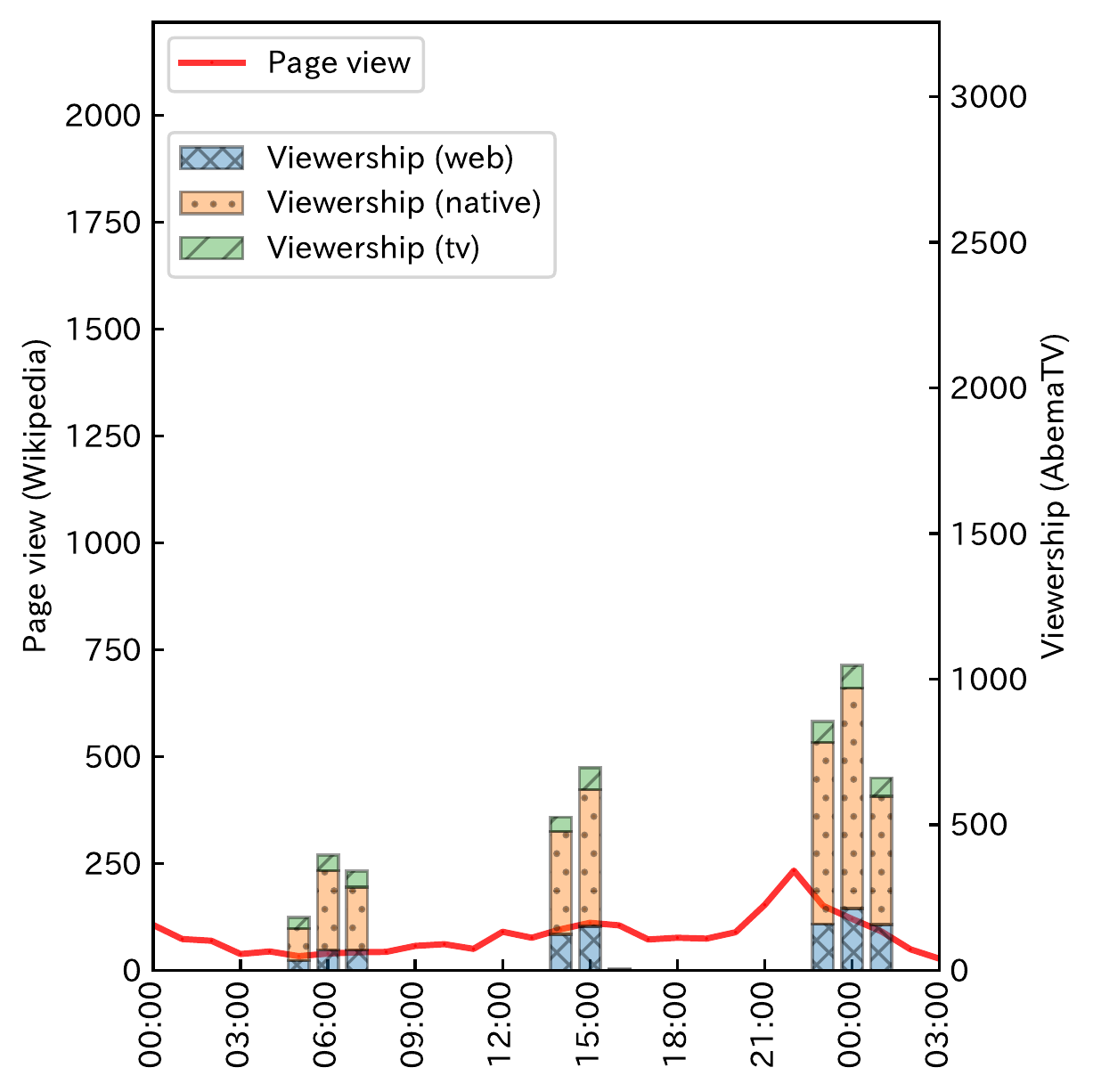}
    \label{fig:article_A1}
  }
  \hfil
  \subfloat[Multiple episodes of one program.]{
    \includegraphics[width=0.48\linewidth]{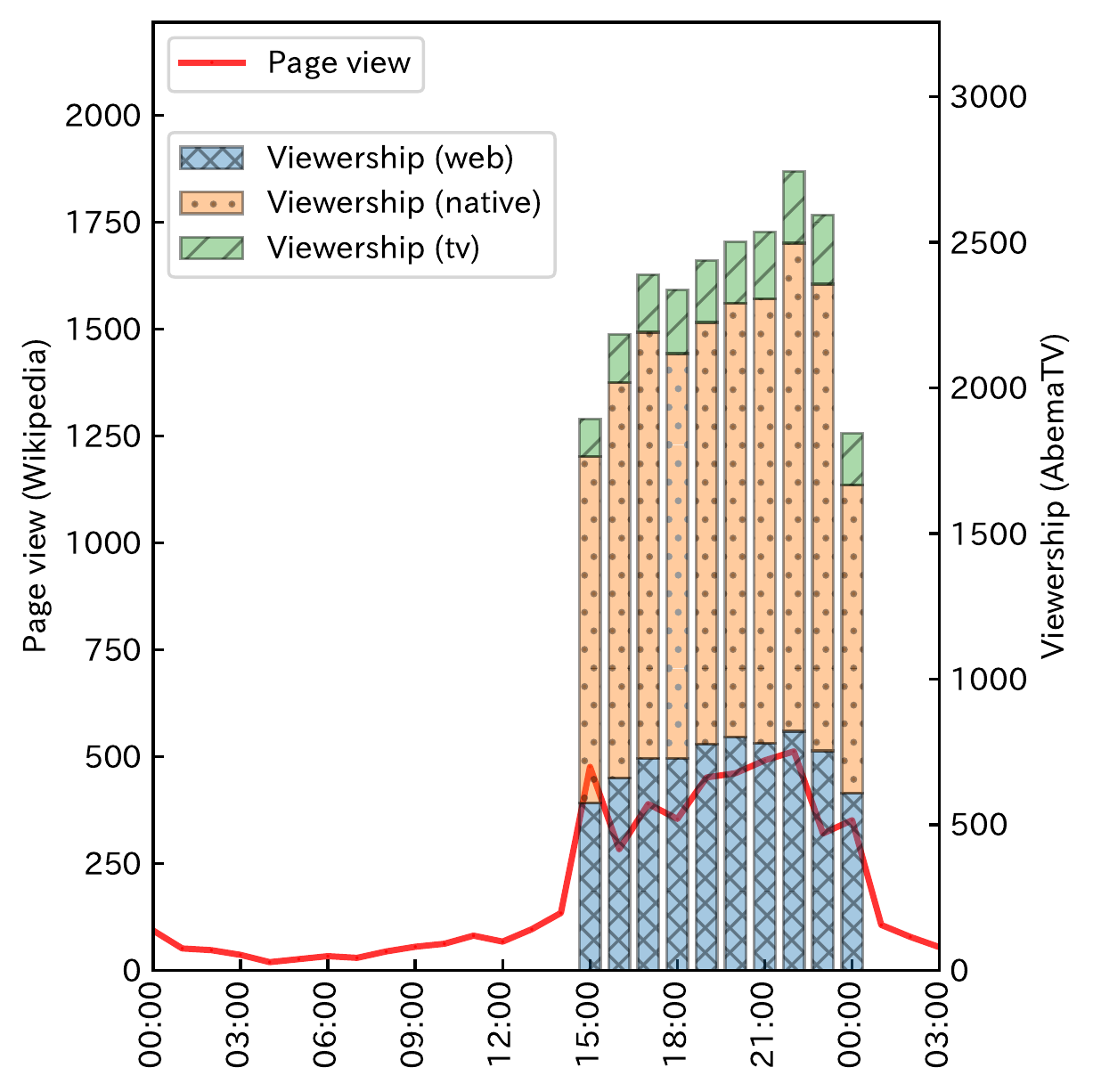}
    \label{fig:article_A2}
  }
  }
  \caption{Viewership and page views of a drama programs: The y-axis is a linear scale. The program of multiple episodes tends to have more page views than broadcasting one episode at a time.}
  \label{fig:article_A}
\end{figure*}

Time series changes of VS and PV of a drama are shown in Fig.~\ref{fig:article_A}. 
Fig.~\ref{fig:article_A1} illustrates a case of broadcasting one episode at a time, and Fig.~\ref{fig:article_A2} shows multiple episodes being broadcast consecutively.
We found that multiple consecutively episodes tend to have higher total associated PV than when one episode is broadcast at a time.

We hypothesized that PV tend to increase with longer broadcasting times such as for programs with multiple episodes.
We examined Pearson correlation coefficients by length of broadcasting time (Airtime), total PV, and total VS of the program.
The Pearson correlation coefficients are shown in TABLE~\ref{tab:corr}.
The results show that the correlation coefficient between Airtime and PV was as low as 0.2.
On the other hand, the correlations between Airtime and VS and between VS and PV are relatively high.
There is no direct relationship between Airtime and PV; 
instead it is assumed that increased Airtime leads to an increase in VS, and this in turn leads to an increase in PV.

\begin{table}[tp]
  \centering
  \caption{Correlation coefficients: The correlation between Airtime and PV were low, but the correlation between Airtime and VS or VS and PV is relatively high.}
\begin{tabular}{ l | r  r  r  r  r }
\hline
  & PV\_pc & PV\_mobile & PV\_SUM & Airtime \\
\hline
Airtime & 0.2098 & 0.1814 & 0.1955 &  \\
VS\_web & 0.4964 & 0.4024 & 0.4446 & 0.4500 \\
VS\_native & 0.4344 & 0.4144 & 0.4308 & 0.5458 \\
VS\_tv & 0.3167 & 0.3025 & 0.3143 & 0.4089 \\
VS\_SUM & 0.4722 & 0.4270 & 0.4526 & 0.5369 \\
\hline
\end{tabular}
  \label{tab:corr}
\end{table}

\subsection{Time until maximun PV is achieved}

As shown in Fig.~\ref{fig:article_A}, PV is not always at maximum during broadcast.
For example, in Fig.~\ref{fig:article_A1}, PV reached maximum just before the third program began.
The point at which PV reaches its maximum can be classified into three times: before, during, and after broadcast.
Here, we define that before broadcast as from 24 hours before the start time to start time itself;
during broadcast is from the start to the end of the broadcast;
and after broadcast is from the end of the broadcast up to 24 hours after it.

\begin{figure}[tp]
  \centering
  \subfloat[All programs]{
    \includegraphics[width=0.99\linewidth]{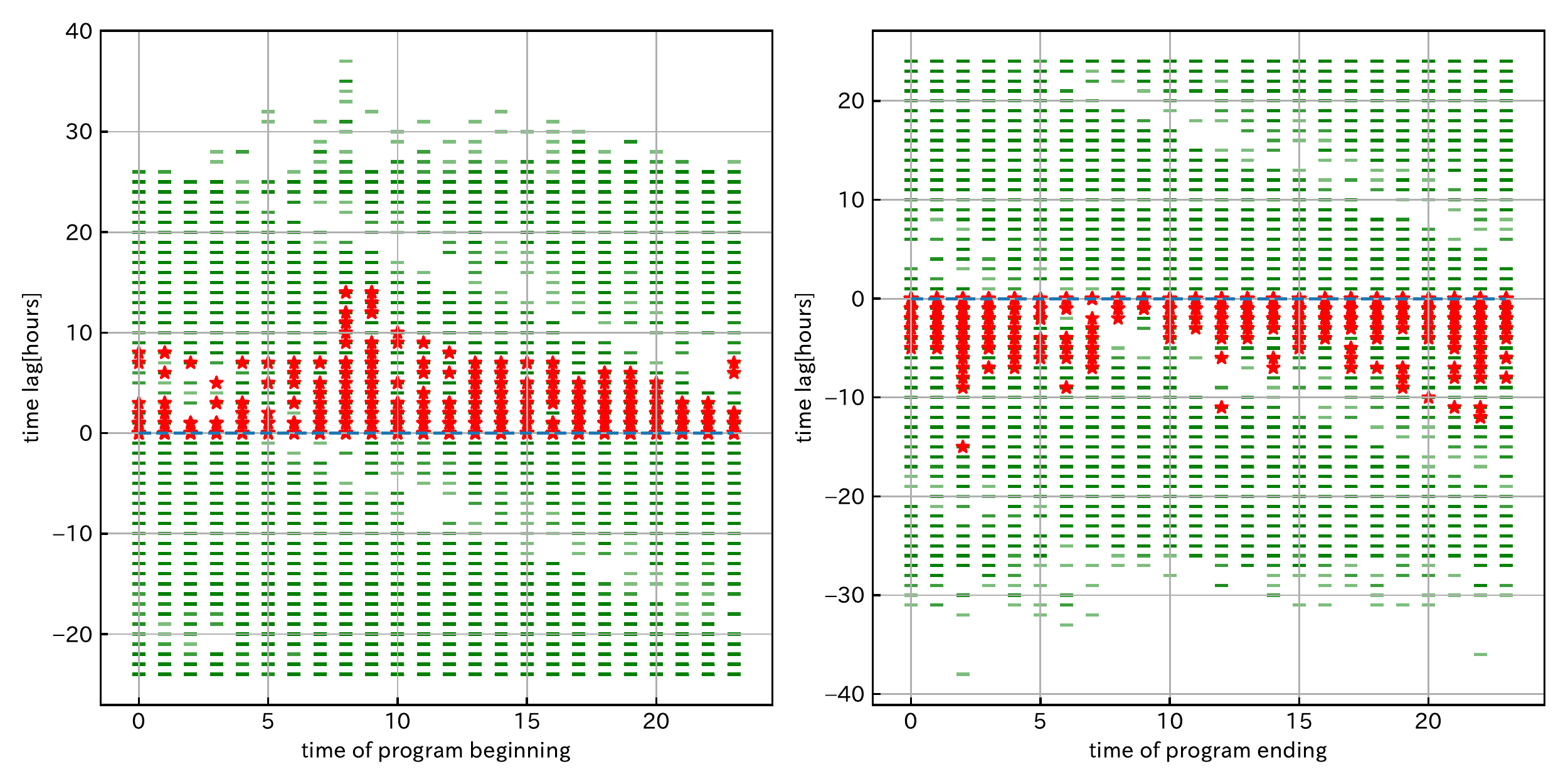}
    \label{fig:time_lag_24_all}
  }
  \hfil
  \subfloat[Animation]{
    \includegraphics[width=0.99\linewidth]{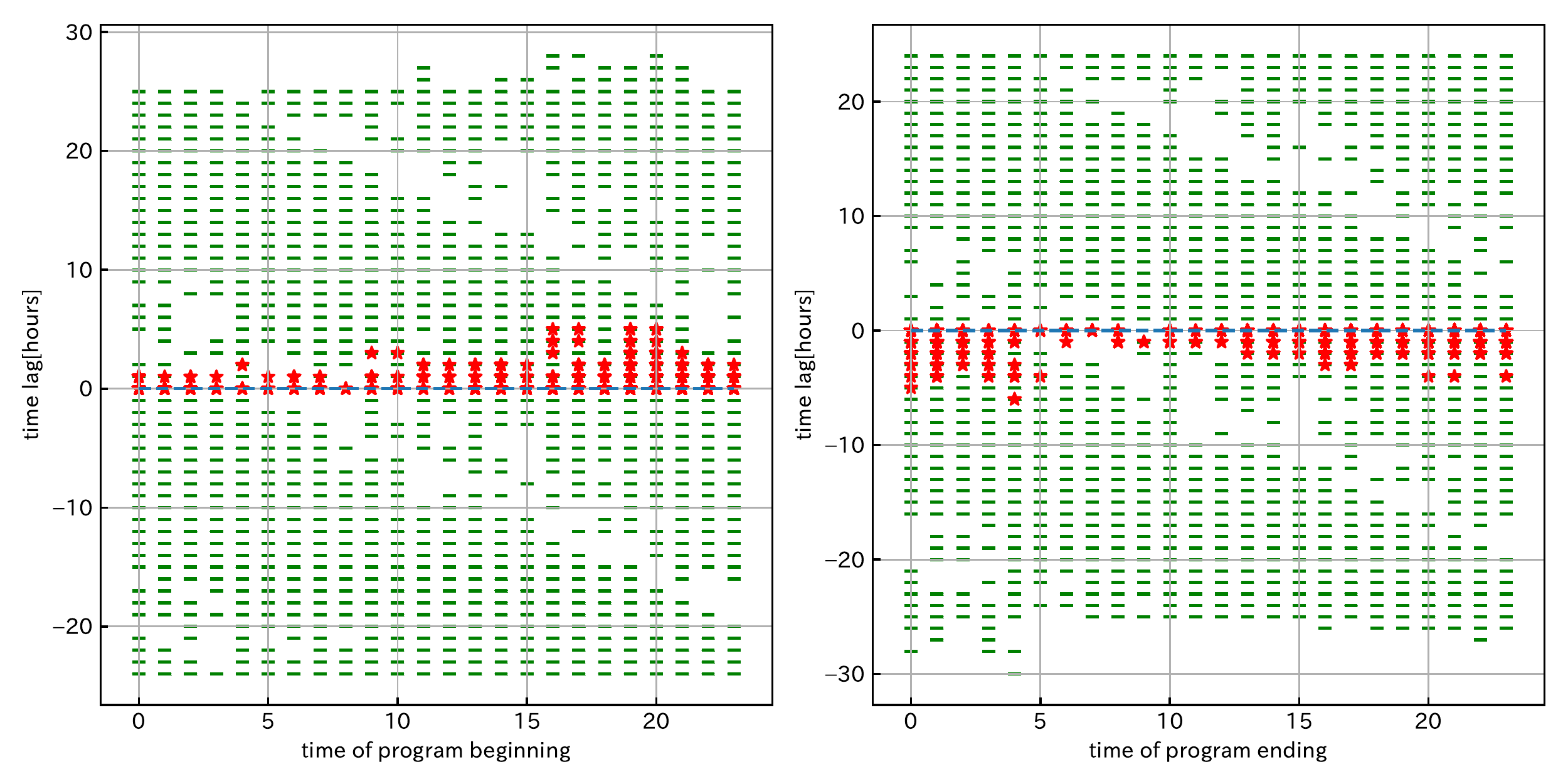}
    \label{fig:time_lag_24_animation}
  }
  \hfil
  \subfloat[Movie]{
    \includegraphics[width=0.99\linewidth]{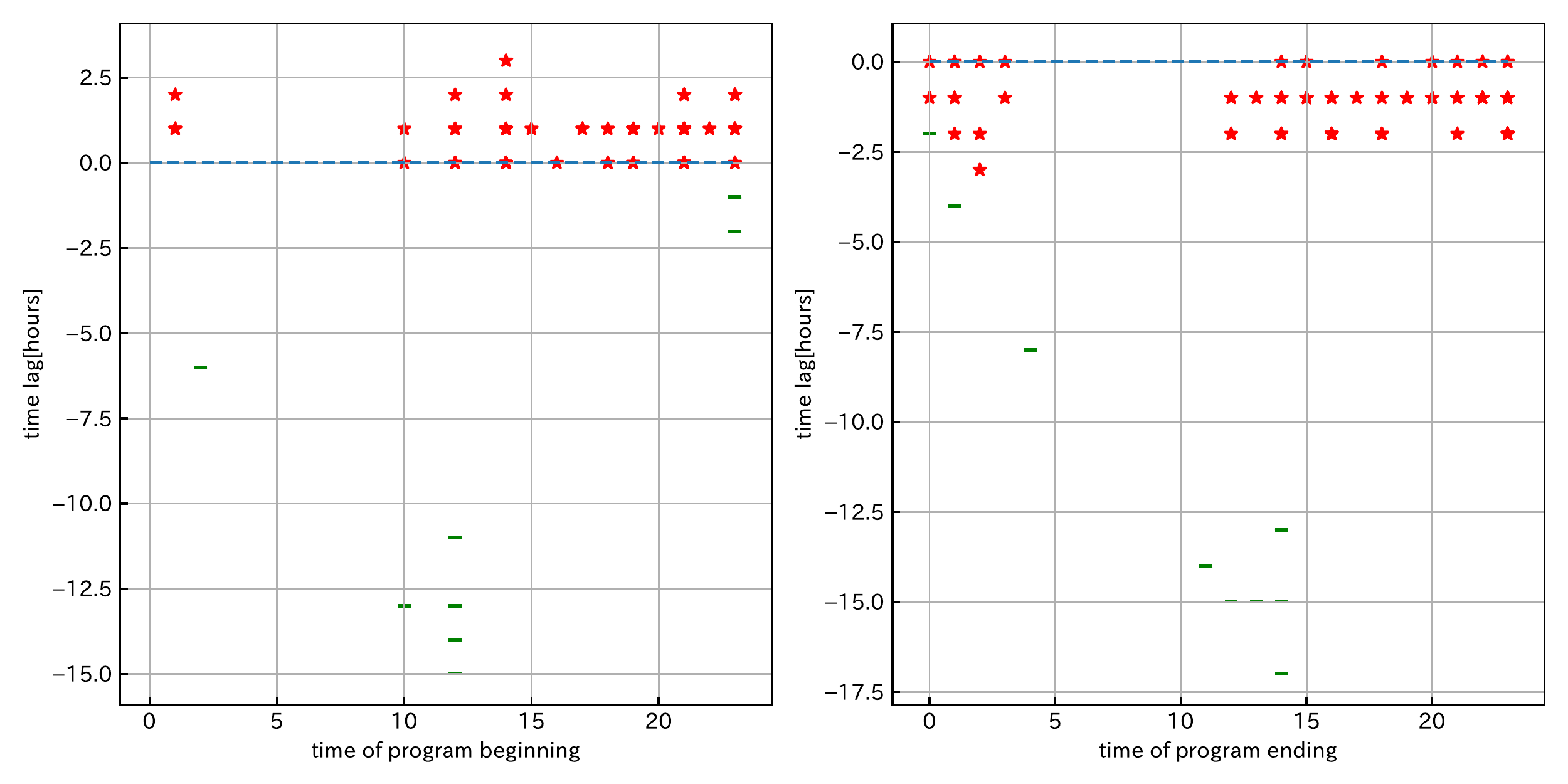}
    \label{fig:time_lag_24_movie}
  }
  \caption{Time until PV is maximized: ``$\star$'' indicates during broadcast and ``$-$'' indicates before or after broadcast. Zero on the y-axis indicates broadcast start time (left) and broadcast end time (right). The time lag is affected by hour in most genres although movies are hardly affected.}
  \label{fig:time_lag_24}
\end{figure}

The lag from broadcast start to maximum PV and the lag from broadcast completion to maximum PV are shown in Fig.~\ref{fig:time_lag_24}.
``$\star$'' indicates maximum PV during broadcast where ``$-$'' indicates it was reached before or after broadcast.
As broadcast start time moves beyond 9:00, the time lag becomes gradually shorter;
after 23:00 PV tends to be maximized during the broadcast itself.
This trend is observed especially in animation (Fig.~\ref{fig:time_lag_24_animation}) 
and is considered a ``prime time'' in which Wikipedia PV is likely to increase.
On the other hand, a similar trend is not observed in movies as maximum PV tends tone reached during the broadcast (Fig.~\ref{fig:time_lag_24_movie}).

\subsection{Maximum PV by time of day}

In the previous section, we investigated the time lag until PV is maximized in relation broadcast time.
Wikipedia articles have mobile and PC pages which are accessed by different devices; 
we now consider and analyze users' access to articles at different times by each device type.

\begin{figure}[tp]
  \centering
  \includegraphics[width=0.99\linewidth]{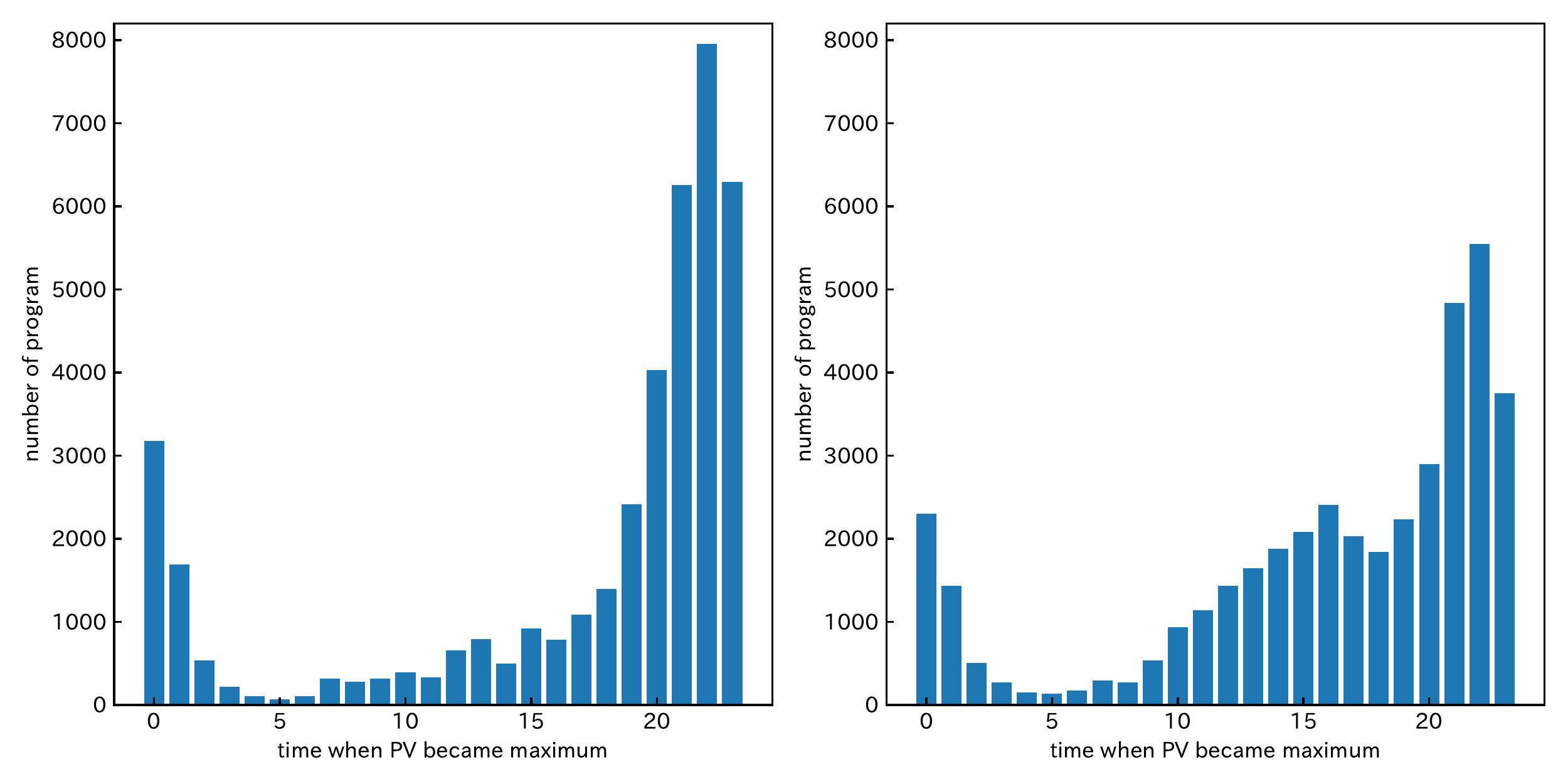}
  \caption{Maximum PV of all programs by time of day. PC pages (right) tend to be more accessible during the daytime than mobile pages (left).}
  \label{fig:count_max_hour}
\end{figure}

The distribution of time of day when mobile and PC PVs is maximized is shown in Fig.~\ref{fig:count_max_hour}.
Both type of PV tend to reach maximum around midnight.
PC pages tend to be more heavily accessed during the day than mobile pages.
The number of programs per hour is almost constant.

\begin{figure}[tp]
  \centering
  \includegraphics[width=0.99\linewidth]{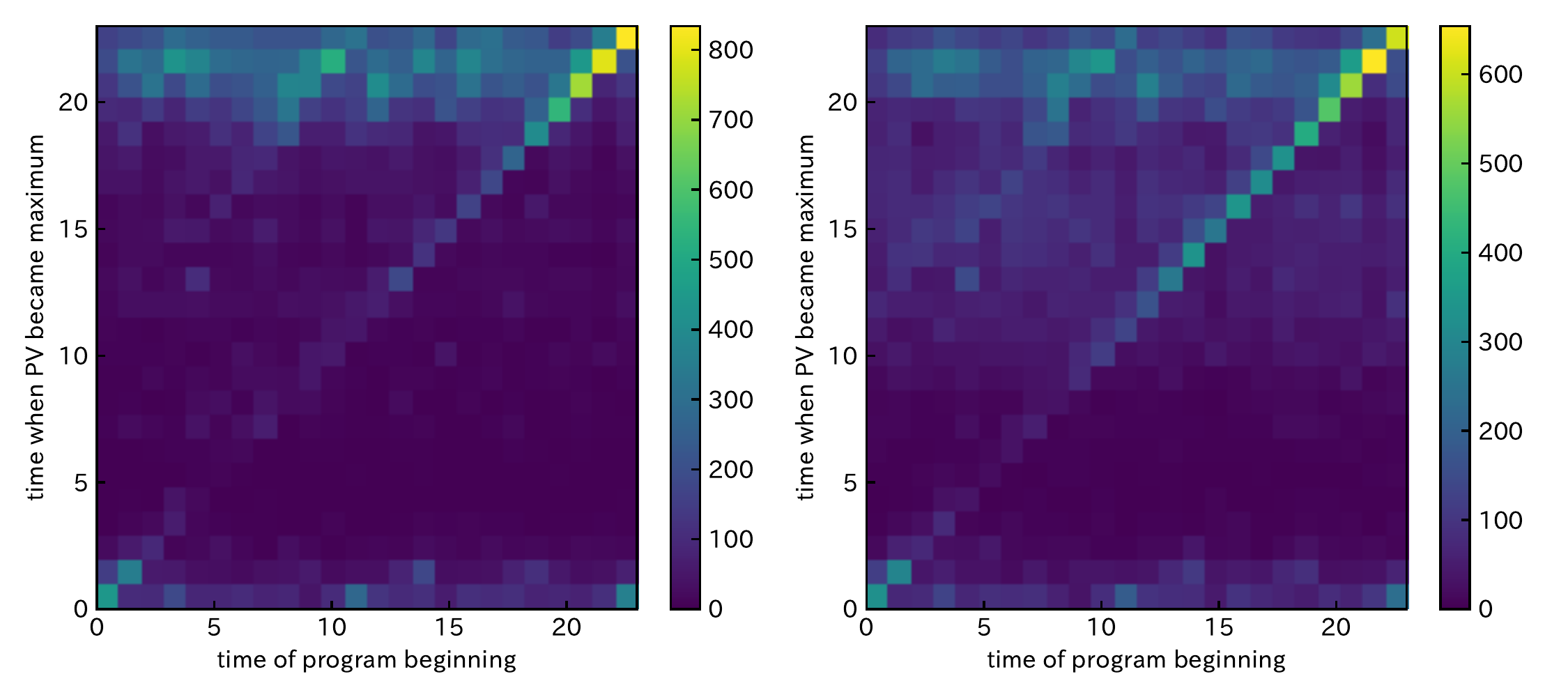}
  \caption{Distribution of the broadcasting start time and the time when the PV becomes the maximum in all programs for mobile (left) and PC (right). Since the diagonal values in the figure tend to be higher, the broadcasting time also has an impact.}
  \label{fig:distribution_allgenre}
\end{figure}

We further observe not only PV but also the relationship between PV and the time of broadcast.
Fig.~\ref{fig:distribution_allgenre} illustrates the distribution of broadcast start time and the point at which maximum PV is achieved for all programs.
The higher diagonal values indicate that time of broadcast has an impact on PV.

\subsection{VS and PV by the number of broadcasts}

In the previous section, we focused on the time of broadcast.
Here, we consider when a program has more than one episode. 
For example, there is an animation series composed of 12 episodes which are broadcast consecutively;
are VS and PV diffrent between the first and last episodes?
We analyze the relationship between the number of broadcasts and the VS or PV.

\begin{figure}[tp]
  \centering
  \subfloat[All programs]{
    \includegraphics[width=0.99\linewidth]{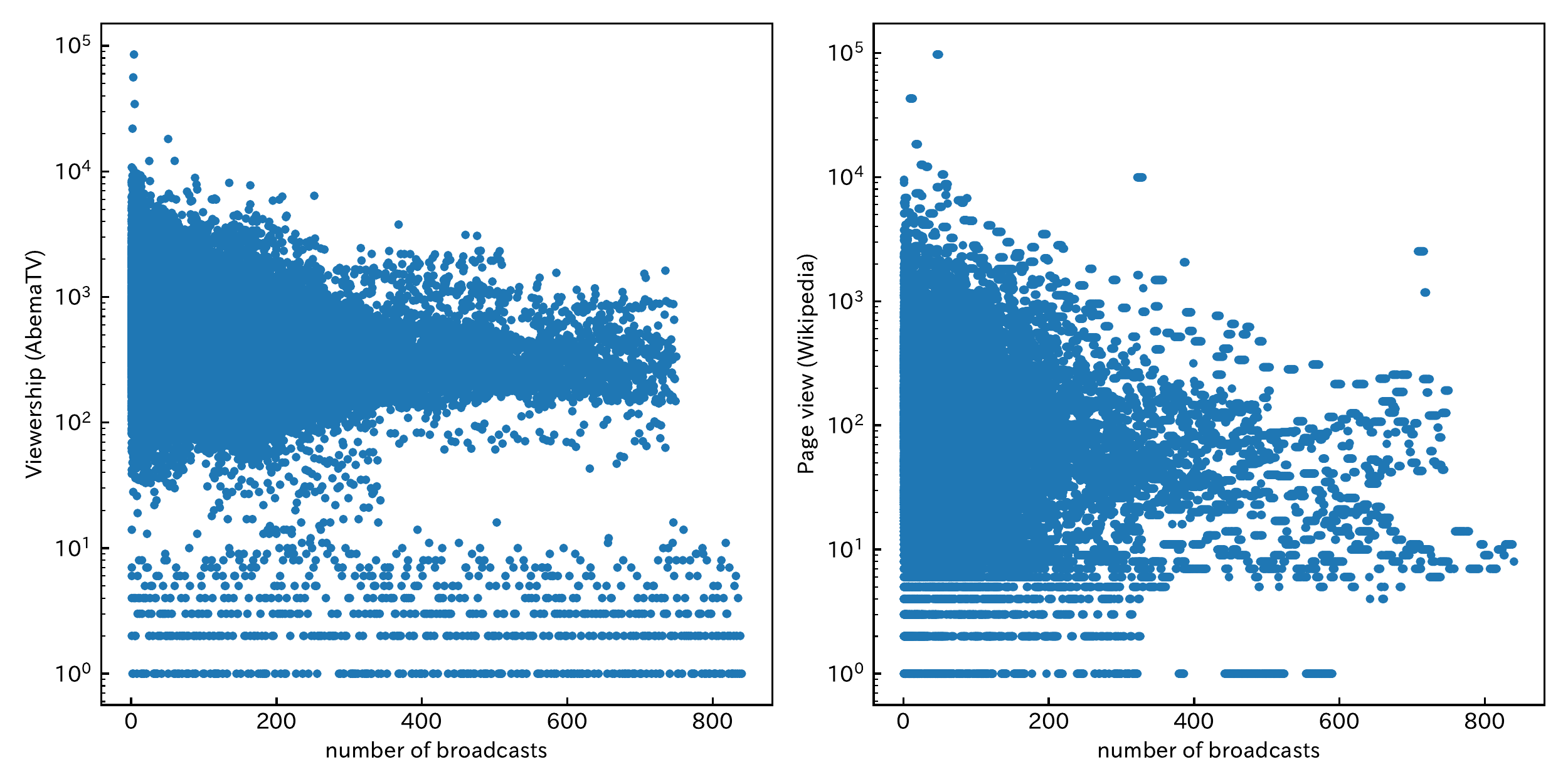}
    \label{broadcast_times-allgenre}
  }
  \hfil
  \subfloat[Animation]{
    \includegraphics[width=0.99\linewidth]{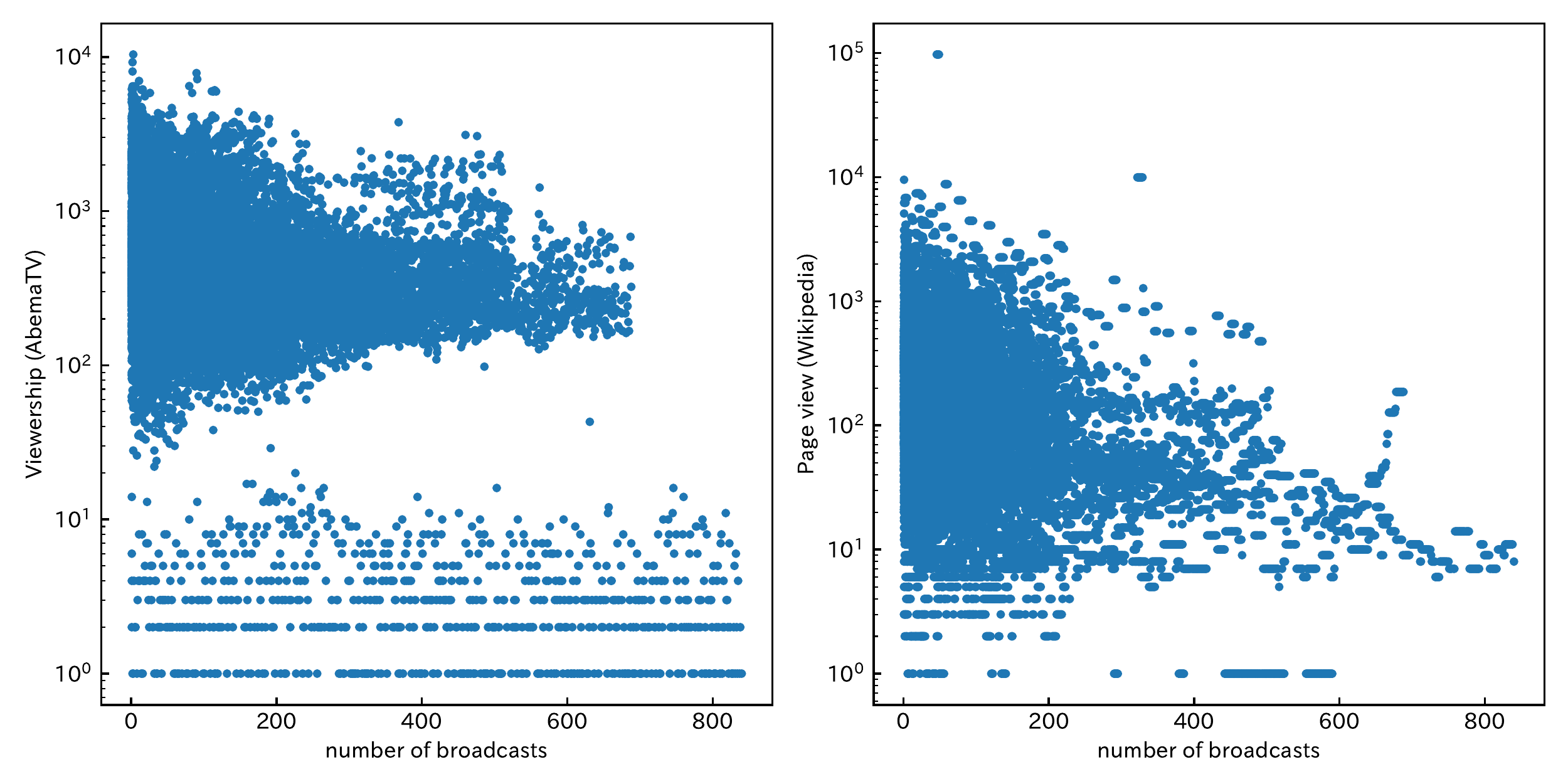}
    \label{broadcast_times-animation}
  }
  \hfil
  \subfloat[News]{
    \includegraphics[width=0.99\linewidth]{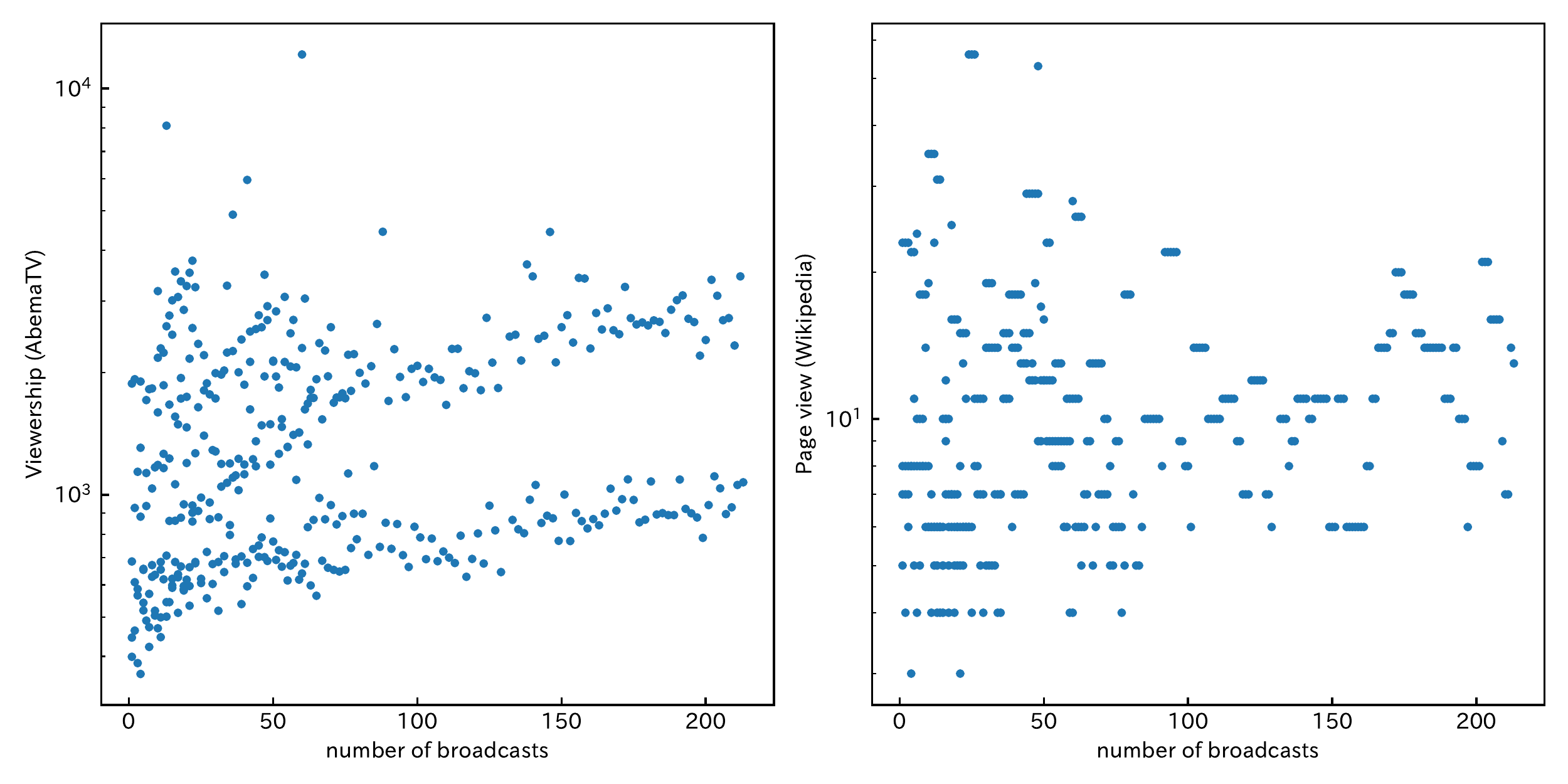}
    \label{broadcast_times-news}
  }
  \hfil
  \subfloat[Soccer]{
    \includegraphics[width=0.99\linewidth]{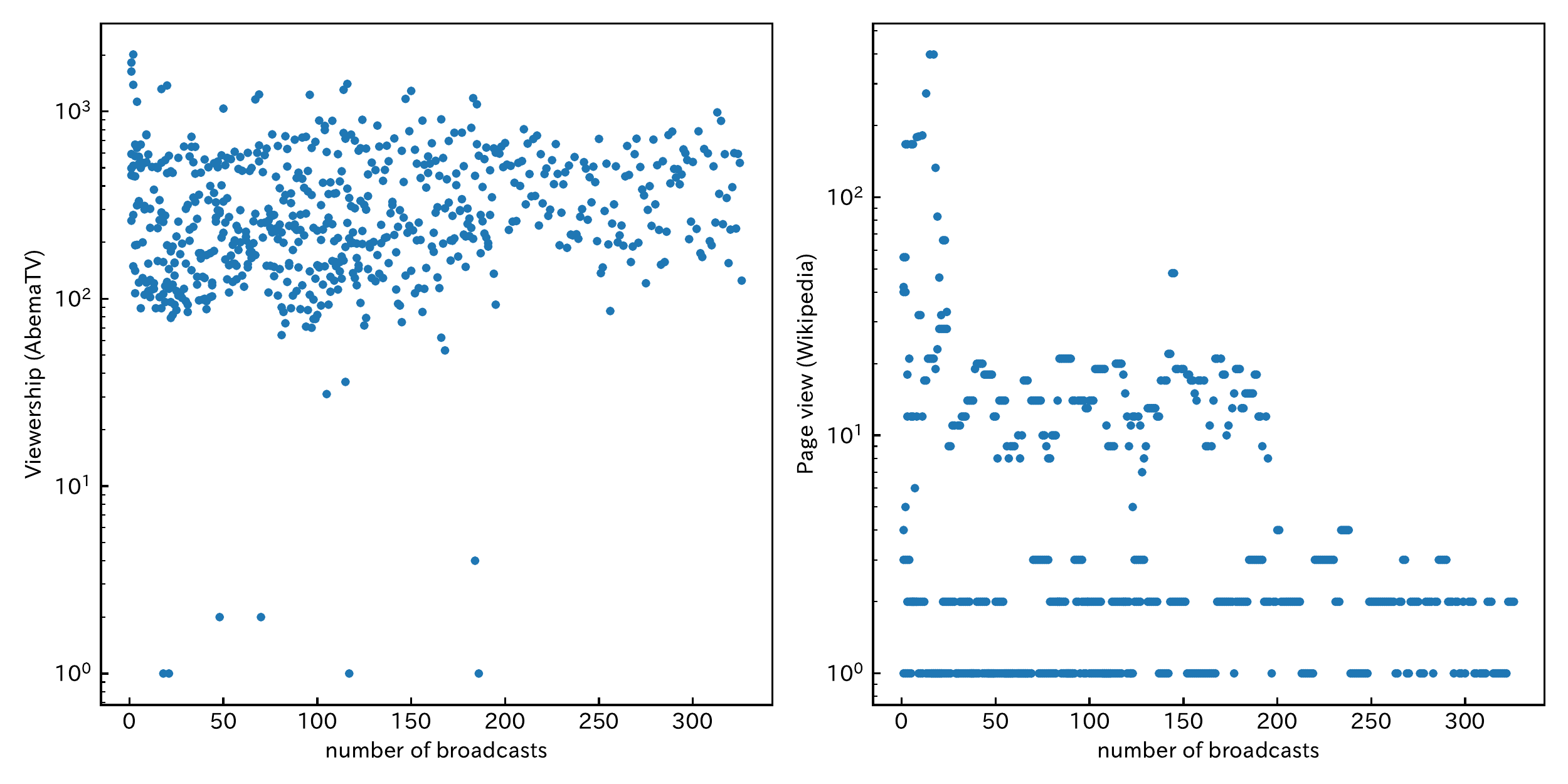}
    \label{broadcast_times-soccer}
  }
  \caption{Scatter diagram of number of broadcasts and VS (left) and PV (right). The y-axis is a logarithmic scale. As the number of broadcasts increases, VS tends to stabilize but PV tends to decreases.}
  \label{fig:broadcast_times}
\end{figure}

Fig.~\ref{fig:broadcast_times} is scatter diagram of the number of broadcasts and VS and VP values;
this is limited to data for the months from June to October 2017.
When an episode of a series appears for the first time in this period, we consider this the first episode of the series.
As the number of broadcasts increases, VS tends to stabilize but PV tends to decrease.
This trend is especially apparent for animation programs.
For news broadcasts, VS tends to increase as the number of broadcasts increases 
while, for soccer programs, VS has a certain tendency regardless of the number of broadcasts.
The PV values relating to news and soccer programs were small and no distinguishing features were identified.

\section{Conclusion}

In this study, we focused on Internet TV station and searching Wikipedia as exploratory behavior on the web.
We analyzed the influence of Internet TV station on Wikipedia page views.
Our aim was to clarify page view characteristics as related to Internet TV station in order to index outward impact and develop a prediction model.

The results indicate that there is a correlation between viewership and page views.
We also found that the time lag between TV and web gradually reduces as the broadcast start time moves beyond 9:00;
after 23:00, page views tend to reach maximum during the broadcast itself.
Moreover, we differentiated between mobile and PC page views 
and found that PC pages tend to be more heavily accessed during the daytime.
We considered the number of broadcasts, 
and we observed that viewership tends to stabilize as the number of broadcasts increases; but that page views tend to decrease.

\bibliographystyle{IEEEtran}
\bibliography{IEEEabrv,references}

\end{document}